\documentclass[
 reprint,
nofootinbib,
 amsmath,amssymb,
 aps,
 prd,
]{revtex4-2}

\usepackage{graphicx}
\usepackage{dcolumn}
\usepackage{bm}
\usepackage{placeins}
\usepackage[normalem]{ulem}
\usepackage[utf8]{inputenc}

\usepackage[usenames]{color}
\usepackage[colorlinks,bookmarks]{hyperref}
\definecolor{linkblue}{rgb}{0,0,0.8}
\definecolor{linkgreen}{rgb}{0,0.5,0}

\hypersetup{pdfpagemode=UseNone, pdfstartview=FitH, linkcolor=linkblue, %
            citecolor=linkgreen, urlcolor=linkblue}

\begin{document}

    \title{Baryonic effects on CMB lensing and neutrino mass constraints} 

\author{Eegene Chung${}^{1,2}$}
\author{Simon Foreman${}^{3,4,1}$}
\author{Alexander van Engelen${}^{5,1}$}
\affiliation{ ${}^1$Canadian Institute for Theoretical Astrophysics, University of Toronto, 60 St.~George St., Toronto, ON, Canada M5S 3H8}
\affiliation{ ${}^2$Department of Physics, University of Toronto, 60 St.~George St., Toronto, ON, Canada M5S 1A7}
\affiliation{ ${}^3$Perimeter Institute for Theoretical Physics, 31 Caroline St.~North, Waterloo, ON, Canada N2L 2Y5}
\affiliation{ ${}^4$Dominion Radio Astrophysical Observatory, Herzberg Astronomy \& Astrophysics Research Centre, National Research Council Canada, P.O.\ Box 248, Penticton, BC, V2A 6J9, Canada}
\affiliation{ ${}^5$School of Earth and Space Exploration, Arizona State University, Tempe, AZ, 85287, USA}


\begin{abstract}
    Measurements of  gravitational lensing of the cosmic microwave background (CMB) hold the promise of yielding unique insights into cosmology at high redshift.  Uncertainties due to baryonic effects associated with galaxy formation and evolution, including gas cooling, star formation, and feedback from active galactic nuclei (AGN) and supernovae, have typically been neglected when forecasting the sensitivity of future CMB surveys.
    In this paper, we determine the impact of  these effects 
    using four suites of hydrodynamical simulations which incorporate various prescriptions for baryonic processes, namely OWLS, BAHAMAS, Horizon, and IllustrisTNG.     Our analysis shows characteristic power suppressions of several percent  in  CMB lensing due to baryonic effects, compared to dark-matter only simulations, at experimentally observable angular scales.   
            We investigate the associated bias in the inferred neutrino mass for experiments like  
    the upcoming Simons Observatory and CMB-S4.  
    Depending on the experimental precision and the strength of the baryonic feedback within the simulations,
    biases in the neutrino mass sum show significant dispersion, ranging from very small to an over-estimation by 1.8$\sigma$.  We conclude that baryonic effects will likely  be non-negligible for a detection of neutrino mass using CMB lensing.
\end{abstract}

\maketitle

\section{\label{sec:intro}Introduction}

Gravitational lensing of the cosmic microwave background (CMB)~\cite{Lewis:2006fu} has emerged as a powerful probe of cosmology. After numerous detections with data from ground-based telescopes (e.g.~\cite{Ade:2013gez,Array:2016afx,Sherwin:2016tyf,Wu:2019hek}) and the WMAP and Planck satellites (e.g.~\cite{Smith:2007rg,Hirata:2008cb,Aghanim:2018oex}), a base set of analysis techniques is now well-established. Future measurements promise to contribute to several scientific goals, such as improving constraints on primordial non-Gaussianity, neutrino mass, and the cosmic growth history~\cite{Schmittfull:2017ffw,Yu:2018tem}; enabling precise estimates of galaxy cluster masses~\cite{Raghunathan:2017cle}; providing a valuable input to primordial gravitational wave searches by allowing for ``de-lensing" of CMB B-mode polarization maps~\cite{Simard:2014aqa}; and improving calibration of cosmic shear measurements at lower redshifts~\cite{Schaan:2016ois}. 

To make good on this promise, however, upcoming lensing measurements will need to control for a plethora of systematics related to beam calibration \cite{Miller:2008zi}, galactic and extragalactic foregrounds \cite{vanEngelen:2013rla,Osborne:2013nna,Ferraro:2017fac}, non-Gaussian clustering of the lenses \cite{Bohm:2016gzt,Bohm:2018omn,Beck:2018wud,Fabbian:2019tik}, and post-Born effects~\cite{Pratten:2016dsm,Fabbian:2017wfp,Bohm:2018omn,Beck:2018wud,Fabbian:2019tik}.  In addition, we must have confidence in our modelling of the lensing signal itself. The most commonly considered statistic, the lensing potential power spectrum, is directly related to a line-of-sight projection of the underlying matter power spectrum, so any modelling uncertainties on the latter will inevitably translate to the former.

Most predictions for the matter power spectrum assume that gravity is the only significant force affecting the distribution of matter on cosmological scales, but the reality is not so simple. At distances smaller than a few megaparsecs or so, the distribution of baryons is affected by processes like gas cooling, star formation, and feedback, and the distribution of dark matter will also be affected, although to a lesser extent. These ``baryonic effects" on clustering have been identified as a key issue for upcoming cosmic shear surveys~\cite{White:2004kv,Zhan:2004wq,Jing:2005gm,Rudd:2007zx,Semboloni:2011fe,natarajan2014}, affecting two-point statistics by several percent, far greater than the precision that will be required of the associated theoretical predictions~\cite{Huterer:2004tr,Hearin:2011bp}. Because of this, a variety of strategies have been proposed to mitigate this uncertainty for the purpose of obtaining cosmological constraints, as summarized recently in Ref.~\cite{Chisari:2019tus}. Alternatively, cosmic shear observations can be seen as an avenue to learn about baryonic effects themselves~\cite{Harnois-Deraps:2014sva,Foreman:2016jzy,Foreman:2019}.

The impact of baryonic effects on CMB lensing, on the other hand, is not nearly as well-studied. Ref.~\cite{natarajan2014} found that a lensing-based neutrino mass constraint would be biased by $\sim10\,{\rm meV}$ if baryonic effects were given by the OWLS-AGN simulation~\cite{Schaye:2009bt,vanDaalen2011} but were ignored in a parameter analysis, for cosmic-variance--limited lensing measurements over half the sky up to $\ell_{\rm max}=2000$. Ref.~\cite{McCarthy:2017csu} investigated the impact of feedback strength in the BAHAMAS simulations~\cite{McCarthy:2016mry} on cross-correlations of CMB lensing with other observations, finding, for example, that a cross spectrum between CMB lensing and cosmic shear maps varies by 5\% at $\ell=1000$. However, Ref.~\cite{McCarthy:2017csu} did not explicitly look at the impact on the CMB lensing auto spectrum on its own, and mainly focused on tensions between different existing datasets. Ref.~\cite{Merkel:2017amt} found that baryonic effects have a mild impact on parameter errorbars derived from CMB lensing, again based on the OWLS simulations.

In anticipation of upcoming CMB observations by the Simons Observatory (SO)~\cite{Ade:2018sbj} and  possible future CMB experiments such as CMB-S4~\cite{Abazajian:2016yjj,Abitbol:2017nao,Abazajian:2019eic}, PICO~\cite{Hanany:2019lle}, and CMB-HD~\cite{Sehgal:2019ewc}, and also in light of the increasing diversity of cosmological hydrodynamical simulations in existence, it is worthwhile exploring the issue of baryonic effects on CMB lensing in greater depth. In this paper, we quantify the impact of these effects on the CMB lensing power spectrum\footnote{The CMB lensing power spectra we have computed for this work are available at \href{https://github.com/sjforeman/cmblensing_baryons}{\url{https://github.com/sjforeman/cmblensing\_baryons}}.}, based on matter power spectra measured from simulations with a variety of numerical schemes and subgrid models for baryons. By comparing the sizes of these effects with expected power spectrum errorbars for different experiments, we find that these effects will not have a huge impact on lensing measurements from an SO-like survey. On the other hand, several simulations predict an impact that will definitively alter measurements made by a CMB-S4-like survey. A high-resolution survey similar to the CMB-HD concept will detect baryonic effects with high significance, and uncertainties in these effects will therefore propagate into lensing-based cosmological constraints. These results are summarized in Fig.~\ref{fig:fig1} and Table~\ref{tab:chi2bary}.

As an example of this propagation of uncertainty, we estimate the amount by which a measurement of neutrino mass with CMB lensing would be biased by  ignoring baryonic effects. We find a wide range of possible biases, in many cases equal to a significant fraction of the expected statistical uncertainty on the neutrino mass sum, indicating that baryonic effects will need to be modelled appropriately to obtain a robust constraint. These results are summarized in Fig.~\ref{fig:fig2}.

{\bf Note Added:} The original version of this work was based on power spectra that were later found to be affected by an error in the calculations. This error has been fixed in the current version, which has slightly altered the quantitative results but has not qualitatively changed the overall conclusions. The journal version of the paper has been updated with an erratum.

\section{\label{sec:level1}CMB lensing power spectrum}

    \subsection{\label{subsec:method}Method}

        The observed gravitational lensing of the CMB is sensitive to the projected matter fluctuations of the universe. The most common way to quantify the lensing effect is via the convergence power spectrum $C_\ell^{\kappa\kappa}$. It is related to the matter power spectrum $P_{\rm m}(k,z)$ by
        \begin{equation}\label{eq:1}
            C_\ell^{\kappa\kappa}=\int_0^{\chi_{\rm CMB}}\frac{W^\kappa(\chi)^2}{\chi^2}P_{\rm m}(k_{\ell}[\chi], z[\chi]) d\chi
        \end{equation}
        where
        \begin{equation}\label{eq:2}
            W^\kappa(\chi)=\frac{3}{2}\Omega_{\rm m} H_0^2 \frac{\chi}{a(\chi)} 
            \left(
            \frac{\chi_{\rm CMB}-\chi}{\chi_{\rm CMB}}
            \right)
        \end{equation}
        as in Ref.~\cite{bleem2012},
        with $a=\frac{1}{1+z}$ and $k_\ell[\chi]=\frac{\ell+0.5}{\chi}$ following the Limber approximation~\cite{Limber:1953,LoVerde:2008re}.

The matter power spectrum, and therefore the convergence power spectrum, will be affected by baryonic processes, but there is currently no analytical theory that can characterize the impact of these processes on large-scale clustering with sufficient accuracy for future surveys.\footnote{There is hope that perturbation theory could provide such an analytical theory on quasi-linear scales, even in the presence of baryonic effects~\cite{Lewandowski:2014rca,Chen:2019cfu}, but the corresponding predictions will still involve free parameters whose values must either be calibrated using simulations or fit to observations.} Ergo, we must resort to simulations that implement large-scale gravitational clustering alongside hydrodynamics and subgrid models of baryon behavior such as gas cooling, star formation, and feedback. By measuring the matter power spectrum from such simulations, we can determine the impact of baryons on the convergence power spectrum via Eqs.~\eqref{eq:1} and~\eqref{eq:2}.

These simulations have made important strides in recent years, and are increasingly successful in reproducing a variety of observations. For example, previous studies have shown that the simulations including feedback from active galactic nuclei (AGN) can provide agreement with current measurements of optical and X-ray observables of galaxy groups at $z=0$, gas fractions in galaxy clusters, and star formation rates (notably solving the problem of ``overcooling", in which there is an overproduction of stars)~\cite{McCarthy:2009kk,McCarthy:2010dn,vanDaalen2011}. However, different simulations are calibrated to different observations, and the resulting measurements can, in turn, differ for quantities that are not calibrated to match---the matter power spectrum being such a quantity. Therefore, in this paper, we use several distinct suites of simulations in order to quantify the range of effects these simulations imply for CMB lensing.

In particular, we consider the following:
\begin{itemize}
\item OWLS-AGN~\cite{Schaye:2009bt}: We use the ``AGN" run from the OWLS simulation suite, which incorporates thermal AGN feedback in addition to radiative heating and cooling, star formation, and supernova feedback via subgrid prescriptions. The OWLS simulations were run with $L_{\rm box}=100h^{-1}\,{\rm Mpc}$ and $2\times512^3$ particles. OWLS-AGN successfully reproduces several observation of galaxy groups at low redshift.
\item BAHAMAS \cite{McCarthy:2016mry,McCarthy:2017csu}: These simulations have similar subgrid models to OWLS-AGN, but further calibrate them against the low-redshift galaxy stellar mass function and gas fractions in groups and clusters. We use three simulations from the BAHAMAS suite: a fiducial version, and versions with stronger (``High-AGN") and weaker (``Low-AGN") AGN feedback, as implemented via different AGN heating temperatures. Each simulation was run with $L_{\rm box}=400h^{-1}\,{\rm Mpc}$ and $2\times1024^3$ particles.
\item Horizon \cite{Dubois:2014lxa,Dubois:2016,horizon}: The Horizon simulations were run using an adaptive mesh refinement scheme, and were calibrated to reproduce low-redshift black hole scaling relations. The Horizon-AGN run we use here, which implements AGN feedback in a quasar (jet) mode at low black hole accretion rates and a thermal mode at high accretion rates,  also matches measurements of the cosmic star formation history and galaxy luminosity functions. The box size was $100h^{-1}\,{\rm Mpc}$ with $1024^3$ dark matter particles.
\item IllustrisTNG~\cite{Pillepich:2017fcc,Springel:2017tpz,Nelson:2017cxy,Naiman:2018,Marinacci:2017wew,Nelson:2018uso}: These moving-mesh simulations implement AGN feedback thermally at high accretion rates and kinetically (imparting momentum to nearby gas particles) at low accretion rates, building on the earlier Illustris runs~\cite{Vogelsberger:2013eka,Vogelsberger:2014kha,Genel:2014lma,Sijacki:2014yfa} but improving the treatment of galactic winds, stellar evolution, and other processes. We use power spectra from two runs: TNG300-1, with $L_{\rm box}=205h^{-1}\,{\rm Mpc}$ and $2\times2500^3$ particles, and TNG100-1, with $L_{\rm box}=75h^{-1}\,{\rm Mpc}$ and $2\times1820^3$ particles.
\end{itemize}

	In each suite, a simulation with baryonic effects included is paired with a dark matter only (DMO) counterpart, which accounts only for the gravitational clustering of dark matter and baryonic matter, and does not include any baryon-specific effects. We incorporate these effects into the $P_{\rm m}(k,z)$ in Eq.~(\ref{eq:1}) as 
        \begin{equation}\label{eq:3}
            P_{\rm m}(k,z) = P_{\rm theory}(k,z) \frac{\widehat P_{\rm bary}(k,z)}{\widehat P_{\rm DMO}(k,z)}
        \end{equation}
        as done, for example, in~\cite{P_ratio}. For each respective suite, $\widehat P_{\rm bary}(k,z)$ is the measurement from the simulation with baryonic effects, while $\widehat P_{\rm DMO}(k,z)$ is from the corresponding DMO simulation.   $P_{\rm theory}(k,z)$ is a theoretical matter power spectrum that accounts only for gravitational clustering of cold dark matter (CDM) and baryons.

        This scheme of taking the ratio of the measurements from baryonic and DMO simulations serves a few purposes. Firstly, the discrete matter power spectra measured from the simulations are jagged due to large sample variance at scales close to the box size, and hence, ill-suited for direct integration in Eq.~\eqref{eq:1}.Since the simulations are generated with the same initial conditions, we expect much of the sample variance (or cosmic variance) to cancel with the ratio $R(k,z) \equiv \frac{\widehat{P}_{\rm bary}(k,z)}{\widehat{P}_{\rm DMO}(k,z)}$.  This ratio thus gives the mere scaling of the DMO power spectrum due to baryonic effects, which, multiplied by a much smoother theoretical power spectrum $P_{\rm theory}(k,z)$, gives a sufficiently smooth function to effectively integrate over. We generate the theoretical matter power spectrum with CAMB\footnote{Code for Anisotropies in the Microwave Background: \href{https://camb.info}{https://camb.info}}~\cite{Lewis:1999bs}, using the modified halo model from~\cite{mead2016} to incorporate nonlinear growth, and using the cosmological parameters from the final Planck results~\cite{Akrami:2018vks}. Other works~\cite{P_ratio,Mummery:2017lcn,vanDaalen:2019pst} have shown that the ratio in Eq.~\eqref{eq:3} generally has only a weak dependence on the assumed cosmological parameters, justifying our parameterization of $P_m(k,z)$.
        
        Secondly, Eq.~\eqref{eq:1} integrates over all $k$ and $z$, but the simulation measurements are made at a finite range of these variables: at most, $k$ is taken from 0.0628 to $514.7 h\,{\rm Mpc}^{-1}$, and $z$ is taken from 0 to 6, as in the case of OWLS. Beyond this range, we set the matter power ratio $R(k>k_{\rm max},z>z_{\rm max})=1$ (i.e., the case where the baryonic and DMO power spectra are equal), where $k_{\rm max}$ and $z_{\rm max}$ are the maximum $k$ and $z$ values of the measurement range. To extract power spectrum values at $k$ and $z$ values between the discrete measurement points, we use a 2D cubic spline interpolation function on the ratio $R(k,z)$ of the simulation measurements. 
        
        Although the integrand in Eq.~\eqref{eq:1} is non-negligible up to $z=10$, at early times ($z\gtrsim 6$), the baryons have not yet had enough impact on the clustering at the wavenumbers $k$ of interest. We generically use values of $k$ reported for each simulation, which in at least some cases, such as OWLS, goes up to the Nyquist frequency of the measurement grid.  However, by varying the $k_{\rm max}$ and $z_{\rm max}$ values beyond which we set $R=1$, we have verified that $C_\ell^{\kappa\kappa}$ depends on these values by much less than 1\% at least up to $\ell\sim 5000$, which are the scales we use to estimate potential biases on a neutrino mass constraint in Sec.~\ref{sec:mnubias}. At larger multipoles, our computed $C_\ell^{\kappa\kappa}$ curves may slightly underestimate baryonic effects (particularly for BAHAMAS, where we only have power spectra up to $z=3$), but neither at a level that would be visible in our plot in Sec.~\ref{subsec:results} nor at a level that would qualitatively change our conclusions about the impact of baryons on those small scales.

    
The analysis of baryonic effects on the lensing power is only meaningful in the context of whether the effects will be observable in experiments or not: that is, whether the associated changes in the lensing power will be comparable to the errorbars from a given experiment. Baryonic effects could provide an interesting signal on their own, but in this paper, we will mainly treat them as a source of errors or biases in future high-precision measurements.
        With this in mind, the uncertainty on the convergence power spectrum at a single $\ell$ is approximately given by~\cite{Knox:1995dq}:
        \begin{equation}\label{eq:sig_ell}
            \sigma_\ell \equiv \sigma(C_\ell^{\kappa\kappa}) \approx \sqrt{\frac{2(C_\ell^{\kappa\kappa} + N_\ell^{\kappa\kappa})^2}{f_{\rm sky}(2\ell+1)}},
        \end{equation}
        where $f_{\rm sky}$ is the observed sky fraction and $N_\ell^{\kappa\kappa}$ is the noise power spectrum of the reconstructed lensing map.
        The errorbar associated with a bin from $\ell_{\rm min}$ to $\ell_{\rm max}$ (neglecting covariances between different multipoles) is then
        \begin{equation}\label{eq:sig_bin}
            \sigma_{\rm bin}=\left[\sum^{\ell<\ell_{\rm max}}_{\ell=\ell_{\rm min}}\left(\frac{1}{\sigma_\ell^2}\right)\right]^{-1/2}.
        \end{equation}
    
         We consider future experiments, which we categorize as Stage III (similar to the upcoming Simons Observatory \cite{Ade:2018sbj}) and Stage IV (similar to the proposed CMB-S4 project \cite{Abazajian:2016yjj,Abitbol:2017nao,Abazajian:2019eic}), as well as the envisioned CMB-HD experiment  \cite{Sehgal:2019ewc}, which can measure the CMB lensing power spectrum at an unprecedented precision due to its significantly increased collecting area. For Stage III and Stage IV, we assume temperature map noise levels of $6\,\mu{\rm K}$-arcmin and $1\,\mu{\rm K}$-arcmin respectively, and Gaussian beams with width 1.4' for both, while for CMB-HD, we use the lensing sensitivity forecasts from Ref.~\cite{Sehgal:2019ewc}, based on Ref.~\cite{nguyen2018}.
        All calculations for the lensing noise $N_\ell^{\kappa\kappa}$ are obtained for the quadratic reconstruction from Ref.~\cite{Hu:2001kj}, including iterated delensing of polarization \cite{Smith:2010gu}. We assume an observed sky fraction $f_{\rm sky}=0.4$ for all experiments.

    \subsection{\label{subsec:results}Results}
    

      \begin{figure*}
            \includegraphics[scale=0.54]{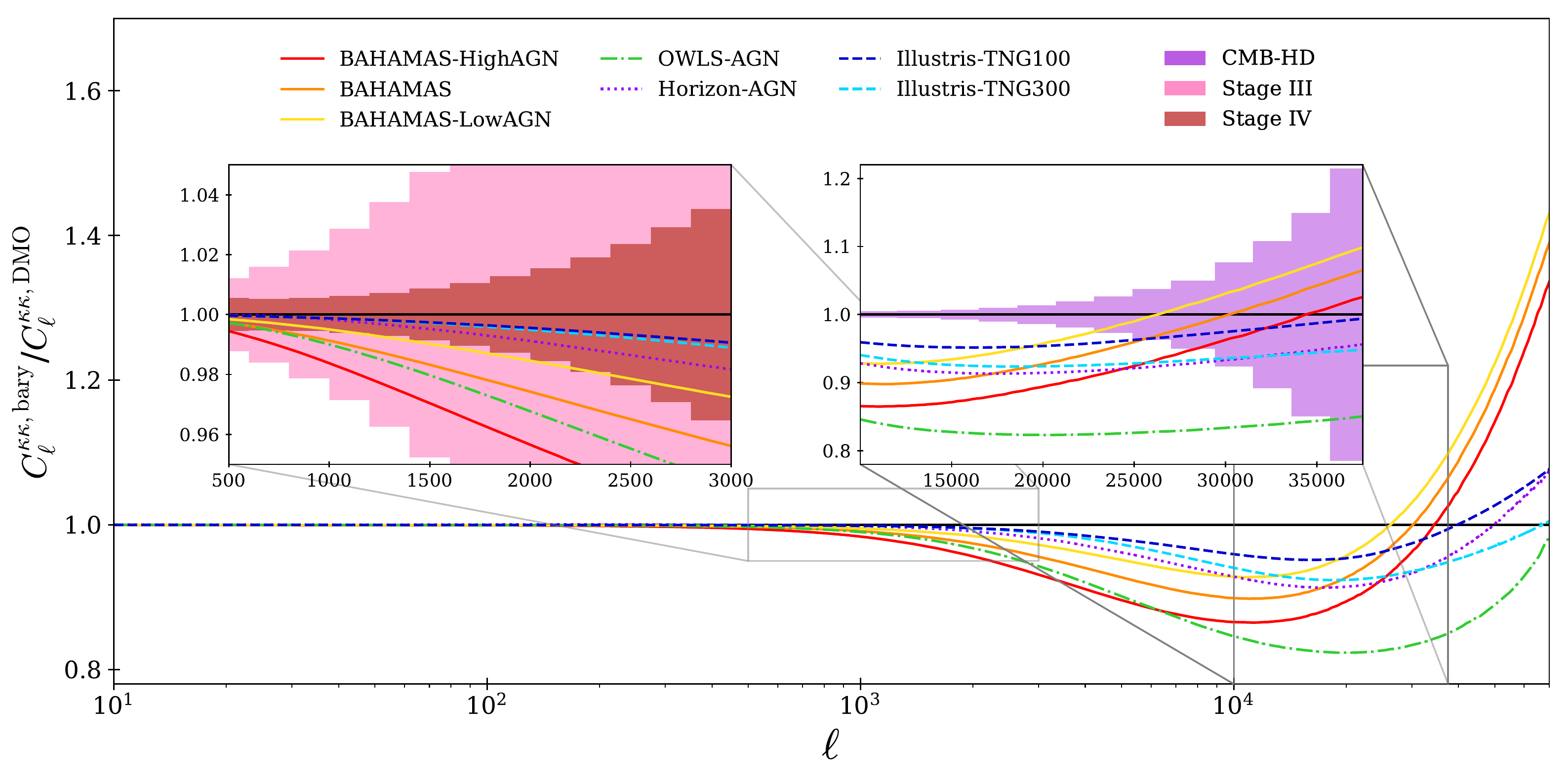}
            \caption{The effect of baryonic feedback on the CMB lensing power spectrum. The curves show the ratio of the baryonic and the dark matter only (DMO) lensing power, and the colors represent the different baryonic simulations used. All curves experience a suppression of varying degree, depending on the prescription of AGN feedback implemented in the simulations. OWLS-AGN (green dash-dotted line) has a maximum suppression of about 17\% compared to its DMO counterpart. Error bars from forecasts of Stage III, Stage IV and CMB-HD experiments are shown in the zoomed-in frames. From $\ell=500$ to $3000$ (left frame), Stage III error bars exceed the effects of baryonic feedback, while Stage IV error bars suggest that the stronger baryonic feedback models will likely be distinguishable from the DMO prediction. In the range accessible to CMB-HD (right frame), the error bars suggest that all baryonic models presented here will be strongly detected.}
            \label{fig:fig1}
        \end{figure*}
        
       \begin{table}
            \renewcommand{\arraystretch}{1.2}
            \begin{ruledtabular}
            \begin{tabular}{l|ccc}
                \textrm{Simulation}  &   \textrm{Stage III}   &   \textrm{Stage IV} &   \textrm{CMB-HD}\\
                \colrule
                OWLS-AGN					&1.2			& 6.8		& 95 \\
                BAHAMAS						&1.0			& 5.6		& 65  \\
                BAHAMAS-LowAGN				&0.60		& 3.3		& 43  \\
                BAHAMAS-HighAGN				&1.8			& 9.7		& 91 \\
                Horizon-AGN					&0.28		& 1.8		& 40  \\
                IllustrisTNG100					&0.16		& 0.94	& 22  \\
                IllustrisTNG300					&0.17		& 1.0		& 32   \\
            \end{tabular}
            \end{ruledtabular}
                \caption{\label{tab:chi2bary}$\sqrt{\Delta \chi^2_{\rm bary}}$ values for each simulation, yielding the effective ``signal to noise" for the impact of baryons, summed over all bandpowers. } 
                            \end{table} 
    
        Fig.~\ref{fig:fig1} shows the ratio of the baryonic lensing power spectra and their DMO counterparts for each suite of simulations, as well as the approximate Stage III, Stage IV, and CMB-HD error bars with a binning scale of $\Delta \ell = \ell_{\mathrm{max}} - \ell_{\mathrm{min}} = 200$. 
        The forecasted errors  provide a measure of  the impact of baryonic feedback  on future surveys of the CMB lensing power spectrum. For, if the suppression due to baryonic feedback is larger than the errors, the measurements will have a source of error that must be accounted for in addition to the statistical error associated with the experiment. 
        
        We find that, as expected, baryons are important at the lower-redshift end of the integral in Eq.~\eqref{eq:1}; for instance, for the Bahamas-HighAGN model the integrands maximally differ at $z = 0.27$ at $\ell = 1000$ for models with and without baryonic effects, with a half-width at half-maximum range of $\Delta z = 0.28$.
        
        To quantify the deviation due to baryons in a single number, which does not depend on the binning scale for the error bars,  we define the effective change in the chi-square statistic that would be obtained if neglecting the impact of baryons:
        	\begin{equation}
        	    \Delta \chi^2_{\mathrm{bary}} = \sum_{\ell} \left ( {{\Delta C_\ell^{\kappa\kappa}} \over {\sigma_\ell} }\right )^2
        	\end{equation}
        	where $\Delta C_\ell^{\kappa\kappa}$ is the difference between the ``true" CMB lensing power spectrum including baryonic effects, and the dark matter only ``fiducial" lensing power spectrum that does not: $\Delta C_\ell^{\kappa\kappa} = C_\ell^{\kappa\kappa}|_{\rm bary} - C_\ell^{\kappa\kappa}|_{\rm DMO}$. This quantity can be considered as the overall ``signal to noise ratio" of the effect of baryons for a given survey.  We show the resulting values for $\sqrt{\Delta \chi^2_{\mathrm{bary}}}$ in Table~\ref{tab:chi2bary}, for each pair of simulation and survey.

        In the range $\ell=500$ to $\ell=3000$, Stage III error bars are so large that baryonic effects from many simulations are nearly insignificant, with a few reaching  $\sqrt{\Delta \chi^2_{\mathrm{bary}}} \sim \mathcal{O}(1)$.  On the other hand, with Stage IV errors, simulations with stronger AGN feedback can become more important, with  $\sqrt{\Delta \chi_{\rm bary}^2}$ as high as $9.7$.

	In the range $\ell=10000$ to $\ell=38000$ where CMB-HD will be the most sensitive, all of the simulations suggest that the baryonic effects due to AGN feedback are clearly observable up to at least $\ell=17000$ (IllustrisTNG300) and at most $\ell=36000$ (OWLS-AGN), beyond which the statistical error blows up. At the angular scales where the precision is highest, different simulations have deviations that vary in their significance, from $\sqrt{\Delta \chi^2_{\mathrm{bary}}}=22$ for IllustrisTNG100 to $\sqrt{\Delta \chi^2_{\mathrm{bary}}}>95$  for OWLS-AGN. The latter shows the highest deviation from its corresponding DMO simulation, with the maximum suppression of around 17\%, whereas IllustrisTNG100 shows the least deviation with maximum suppression around 7\%. In general, the baryonic simulations adopting the AGN feedback model experience a suppression at experimentally observable ranges, and the maximum suppression is reached at varying values of $\ell$ within this range.  After this point, corresponding to the high $\ell$'s and $k$'s is a short distance scale in which galaxy formation takes place. This increases the clustering of matter, causing the lensing power to blow up drastically; however, simultaneously, the statistical errors also blow up, making this observation at the high-$\ell$ range out of reach. In summary, for CMB lensing surveys at very high angular resolution, even the most conservative effects of baryonic feedback are likely to be detectable, and these effects must carefully be accounted for in the systematic error budget. 

\section{\label{sec:mnubias}Neutrino Mass Bias}

    A nonzero neutrino mass results in a damping of the matter power spectrum  due to neutrino free streaming, which occurs because neutrinos are still relativistic when they freeze out of equilibrium with other species \cite{Lesgourgues:2012uu}. The suppression in the matter power brought on by baryonic physics, such as AGN feedback, can resemble the damping introduced by the neutrino masses. If high-precision experiments are sensitive to baryonic effects in CMB lensing (and correspondingly the matter power) measurements at small scales, these effects will bias constraints on the neutrino mass sum. In other words, by underestimating the baryonic effects (that contribute to the damping of the matter power spectrum, similar to neutrinos), we overestimate the neutrino mass sum.
    
    The estimated errors on the set of cosmological parameters $\vec{\theta}=\{\Omega_bh^2, \Omega_ch^2, h, 10^9A_s, n_s, \tau, M_\nu\}$ can be forecasted by the Fisher matrix formalism as in \cite{natarajan2014}:
    \begin{equation}\label{eq:mnubias}
        \Delta\vec{\theta}={\bf F}^{-1}\sum_\ell{\frac{\partial C_\ell^{\kappa\kappa}}{\partial \vec{\theta}}
        \frac{1}{\sigma_\ell^2}\Delta{C_\ell^{\kappa\kappa}}}
    \end{equation}
where $\sigma_\ell$ is given by Eq.~(\ref{eq:sig_ell}) and $\bf{F}$ is the Fisher matrix,
    \begin{equation}\label{eq:fisher}
    {\rm \bf F} = {\rm {\bf C}_{prior}^{-1}} + \sum_\ell{\frac{\partial C_\ell^{\kappa\kappa}}{\partial \vec{\theta}} \frac{1}{\sigma_\ell^2} \frac{\partial C_\ell^{\kappa\kappa}}{\partial\vec{\theta}}},
    \end{equation} 
    with priors incorporated with the  matrix~${\bf C}_{\rm prior}$. Using this formalism, we determine the biases that arise from ignoring baryonic effects: we appropriately set $C_\ell^{\kappa\kappa}|_{\rm true} = C_\ell^{\kappa\kappa}|_{\rm bary}$, i.e. the ``true" CMB lensing includes baryonic effects, and $C_\ell^{\kappa\kappa}|_{\rm fid} = C_\ell^{\kappa\kappa}|_{\rm DMO}$, i.e. where the ``fiducial" lensing does not have baryonic effects. Similarly, taking the neutrino mass sum $M_\nu$ component of Eq.~(\ref{eq:mnubias}) gives the difference between the true neutrino mass sum estimate and the fiducial one. 
     
     For each simulation, we use Eqs.~\eqref{eq:1} to~\eqref{eq:3} to compute $C_\ell^{\kappa\kappa}$ with and without baryonic effects (setting $R(k,z)=1$ for the latter), and inserting them into the above expressions.   We take the covariance matrix between parameters obtained from other measurements, ${\rm {\bf C}_{prior}}$, to have contributions from three sources.  Given that the inference of neutrino mass from upcoming CMB surveys will be limited by the uncertainty in the optical depth due to reionization $\tau$~\cite{Allison:2015qca}, we include two choices of prior: $\sigma_{\mathrm{prior}}(\tau)=0.006$, corresponding to the Planck design sensitivity, and $0.002$, corresponding to the cosmic variance limit.   The  value of $\sigma_{\mathrm{prior}}(\tau)=0.006$ is close to the final official Planck team constraint of $\sigma(\tau)=0.007$~\cite{Akrami:2018vks} and equal to that obtained in a subsequent reanalysis~\cite{Pagano:2019tci}.    The cosmic variance-limited measurement might be achieved with a future ground-based~\cite{Watts:2018etg} or space-based \cite{Matsumura:2016sri,Hanany:2019lle} CMB survey and would depend in detail on our ability to clean large-scale Galactic foregrounds \cite{Errard:2015cxa,Thorne:2019mrd,Remazeilles:2017szm} .   We also include an expansion history prior from baryon acoustic oscillation measurements expected from the DESI survey \cite{Aghamousa:2016zmz}, as well as a prior from the primary CMB temperature and $E$-mode polarization power spectra above $\ell = 300$ from the CMB experiment under consideration.  For this last prior we neglect covariance between this and the CMB lensing measure \cite{Peloton:2016kbw}.
     We use the public code\footnote{\href{https://github.com/msyriac/pyfisher}{https://github.com/msyriac/pyfisher}} from Ref.~\cite{Li:2018zdm} for these computations. 
     
\begin{figure}[t]
        \includegraphics[width = \columnwidth]{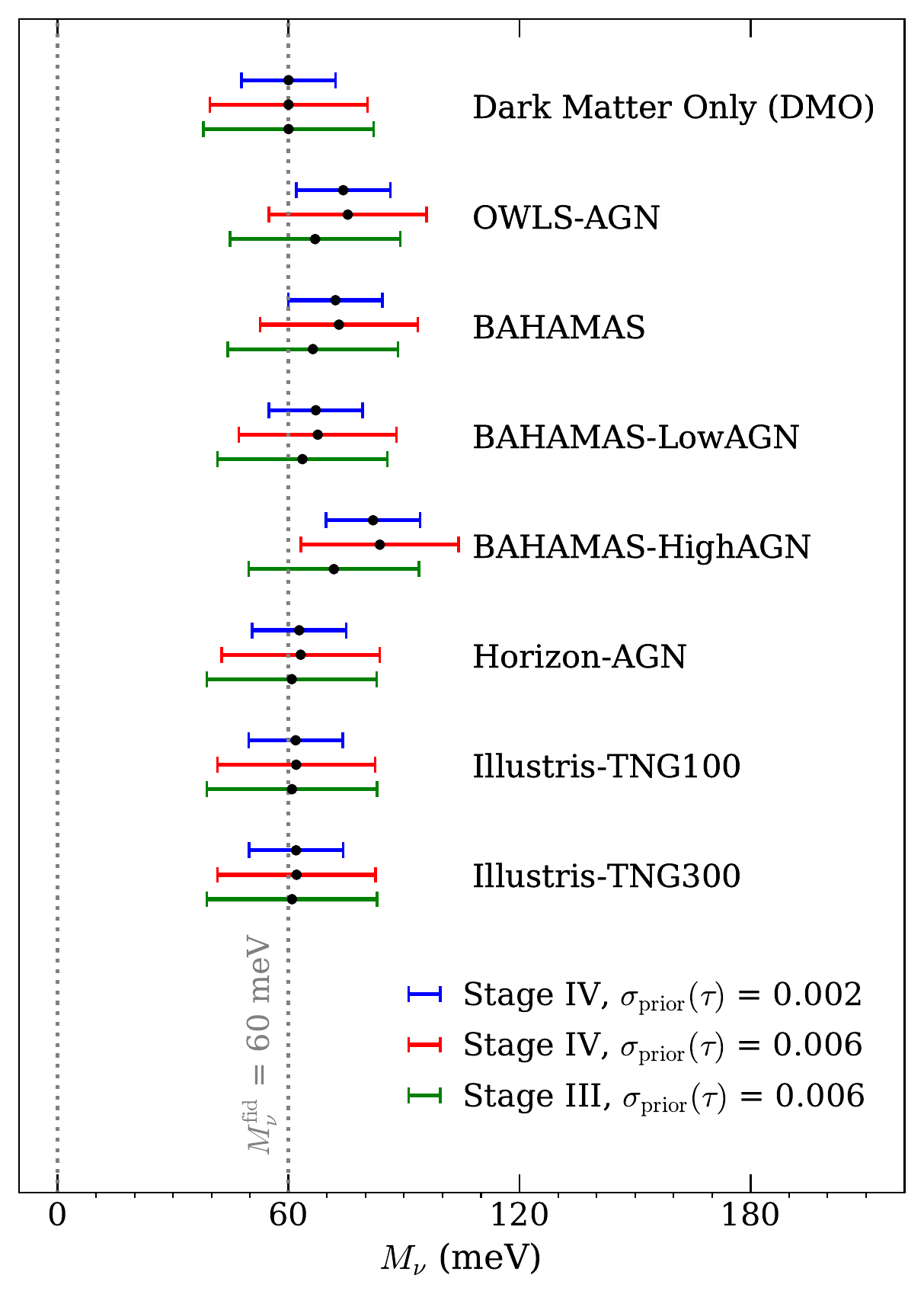}
        \caption{\label{fig:fig2} Neutrino mass bias due to baryonic feedback in large scale clustering, shown as deviations from the fiducial neutrino mass of $M_\nu^{\rm fid}=60$ meV.  Note that for the inverted hierarchy the minimal mass is twice this, at $M_\nu=120$ meV.  We use two values for the uncertainty on the optical depth to reionization, $\sigma_\mathrm{prior}(\tau)$, corresponding to the current constraint of 0.006, and the cosmic-variance-limited constraint of 0.002.  For the Stage III experiment, the bias can be as large as $0.5\sigma$ for the simulation with the strongest AGN feedback.  For Stage IV, the bias is appreciable for many of the simulations, in some cases exceeding~$1\sigma$.   Numerical values are available in Table~\ref{tab:table2}.}
    \end{figure}

     Our estimates for the bias on the neutrino mass sum due to various prescriptions for baryonic feedback are illustrated in Fig.~\ref{fig:fig2}. It is clear that a wide range of bias is present, depending on the different simulations with their own characteristic strength of baryonic effects, and in particular, AGN feedback. Notably, the larger the effect of AGN feedback, the larger the deviation from the fiducial neutrino mass sum ($M_\nu^{\rm fid}$) of 60 meV.  However, the biases are mitigated somewhat by the slightly different shape dependence of the effect of baryons (as shown in Fig.~\ref{fig:fig1}) as compared to massive neutrinos, which give a power suppression that is nearly independent of scale for the range measured by CMB surveys. The tabulated numerical values of these biases can be found in Table~\ref{tab:table2} in Appendix~\ref{app:appB}.  
    
\section{\label{sec:level1}Discussion}

    While negligible for previous measurements, it has become increasingly important to quantify baryonic effects such as star formation, gas cooling, and most eminently, feedback from AGN, on the CMB lensing power spectrum. To study such effects, we have considered a selection of recent suites of hydrodynamical simulations: OWLS, BAHAMAS, Horizon, and IllustrisTNG. Each suite contains pairs of simulations that include or ignore baryonic effects, allowing the impact of baryons to be isolated from clustering due to gravity alone. We have taken matter power spectra measured from each simulation suite and computed the corresponding baryonic effects on the lensing convergence power spectrum, which is related to the matter power spectrum via a weighted line-of-sight projection.

We have shown that the inclusion of these effects in the convergence power spectrum leads to a scale-dependent suppression on the scales that Stage III and Stage IV CMB surveys will probe, with a strength that depends strongly on the specific simulation we consider. Stage~III lensing measurements will reach at most ${\rm S/N}\sim \mathcal{O}(1)$ on these effects, but Stage IV measurements could possibly afford a much stronger detection if feedback is accurately represented by the OWLS or BAHAMAS simulations, while these effects will still have a small impact if given by Horizon or IllustrisTNG (see Fig.~\ref{fig:fig1} and Table~\ref{tab:chi2bary}). The CMB-HD experiment concept, which could measure CMB lensing up to $\ell_{\rm max}\sim 35000$, would see these effects at very high significance if they are represented by any of the simulations we have considered.
    
    Furthermore, the error forecasts from Stage III and Stage IV experiments have allowed us to estimate the bias on the neutrino mass sum that these baryonic effects will incur if not properly accounted for.  For the OWLS-AGN simulation, also studied by Ref.~\cite{natarajan2014}, we found a comparable absolute neutrino mass bias as that reference, but a significantly larger relative bias.  This is due to the tighter statistical uncertainty on $M_\nu$ that we obtained for a Stage IV survey, due to our inclusion of priors from external datasets.  In general, we found that the simulations with higher prescribed AGN effects (OWLS and BAHAMAS) result in biases that are a large fraction of the 1$\sigma$ statistical error, ranging from $0.3\sigma$ to $\sim2\sigma$. Others (Horizon and IllustrisTNG) showed no significant bias on the fiducial neutrino mass sum of 60~meV (see Fig.~\ref{fig:fig2}). The large dispersion between these results indicates that the uncertainties due to baryonic physics will constitute a non-negligible systematic effect for planned CMB surveys, not only for neutrino mass constraints but also for other physics that might leave signatures on similar scales, such as warm~\cite{Bode:2000gq} or fuzzy~\cite{Hu:2000ke,Hui:2016ltb} dark matter.
    
    We note here that we have performed our lensing forecasts using the standard quadratic estimator~\cite{Hu:2001kj}, but that this estimator is known to be sub-optimal on sufficiently small scales. Other lensing reconstruction techniques (e.g.~\cite{Hadzhiyska:2019cle}) can result in lower noise on these scales; use of these techniques would further exacerbate the need to model baryonic effects, since they would be detectable at higher significance in that case.
    
    Luckily, there exist a number of possible methods for controlling for this uncertainty in CMB lensing measurements. The baryon fractions in group-sized halos have been shown to correlate strongly with suppression of the matter power spectrum induced by baryonic effects~\cite{vanDaalen:2019pst}, and these fractions can themselves be probed using secondary effects on the CMB, namely the kinetic~\cite{Battaglia:2016xbi,Battaglia:2017neq}  and thermal~\cite{Ma:2014dea,Tanimura:2019whh,Pandey:2019uxy} Sunyaev-Zel'dovich effects. Thus, for a given CMB survey, there is the potential of calibrating matter power spectrum suppression using data from the very same survey, together with a tracer of large-scale structure. 
    
    Finally, it is possible that results from optical lensing surveys can be used to calibrate the impact of baryons, since the same redshift range is responsible for the effect at both wavelengths.
    
    Other approaches have also been explored for optical lensing of galaxies, including modified halo models~\cite{Semboloni:2012yh,Mohammed:2014mba,P_ratio,Debackere:2019cec}, parametric schemes for modifying $N$-body simulation outputs~\cite{Schneider:2015wta,Schneider:2018pfw}, principal component decompositions of simulation power spectra~\cite{Eifler:2014iva,Huang:2018wpy}, or simple scale cuts to remove data most strongly influenced by these effects (e.g.~\cite{Troxel:2017xyo}). Given the similarities between CMB and optical weak lensing, our results suggest that these approaches should be adopted for CMB lensing as well.  Doing so will be an important step towards unlocking the full potential of CMB lensing to reveal new physics.


\begin{acknowledgments}
We thank Fiona McCarthy for useful discussions, and Marcel van Daalen for providing BAHAMAS power spectra and Francisco Villaescusa-Navarro for measuring and providing IllustrisTNG power spectra used in this work. We also thank the OWLS and Horizon-AGN teams for making their power spectra publicly available, and Mat Madhavacheril  both for making his forecasting code publicly available and for general discussion. We thank Nam Nguyen and Neelima Sehgal for the CMB-HD error bars from Ref.~\cite{Sehgal:2019ewc}.  We thank John Dubinski for patient and essential computing support. E.~C.~was supported by an Undergraduate Student Research Award from the Natural Sciences and Engineering Research Council, held at the Canadian Institute for Theoretical Astrophysics.  Research at the Perimeter Institute is supported in part by the Government of Canada through the Department of Innovation, Science and Economic Development Canada and by the Province of Ontario through the Ministry of Colleges and Universities.
\end{acknowledgments}
\appendix
    \section{\label{app:appB}Neutrino mass bias values}
    Table~\ref{tab:table2} shows the numerical values of the estimated neutrino mass sum bias, corresponding to the different colored bars in Fig.~\ref{fig:fig2}, along with the ratio of the bias and expected $1\sigma$ neutrino mass sum constraint for each simulation-experiment pair. This table also shows the expected $1\sigma$ constraint on its own, for each experiment and for two choices of $\tau$ prior for Stage IV.
    
        \begin{table*}
            \renewcommand{\arraystretch}{1.3} 
            \begin{ruledtabular}
            \begin{tabular}{ccc|lcc}
                Experiment  & $\sigma_{\rm prior}(\tau)$ & $\sigma_{M_\nu}$ [meV] & Simulation & $\Delta M_\nu$ [meV] & $\Delta M_\nu$/$\sigma_{M_\nu}$ \\
                \hline 
                Stage III   &   0.006   &    22     & OWLS-AGN          &  6.9    &   0.31  \\
                            &           &           & BAHAMAS           &  6.3    &   0.29  \\
                            &           &           & BAHAMAS-LowAGN    &  3.6    &   0.16  \\
                            &           &           & BAHAMAS-HighAGN   &  12    &   0.53   \\
                            &           &           & Horizon-AGN       &  0.84 &   0.038  \\
                            &           &           & IllustrisTNG100  &  0.89   &   0.040  \\
                            &           &           & IllustrisTNG300  &  0.90  &   0.041 \\
                \hline 
                Stage IV    &   0.006   &   20      & OWLS-AGN          &  15    &   0.75  \\
                            &           &           & BAHAMAS           &  13    &   0.64  \\
                            &           &           & BAHAMAS-LowAGN    &  7.6    &   0.37  \\
                            &           &           & BAHAMAS-HighAGN   &  24     &   1.2  \\
                            &           &           & Horizon-AGN       &  3.2   &   0.15  \\
                            &           &           & IllustrisTNG100  &  2.0   &   0.096  \\
                            &           &           & IllustrisTNG300  &  2.1   &   0.10 \\
                \hline
                Stage IV    &   0.002   &   12      & OWLS-AGN          &  14    &   1.2  \\
                            &           &           & BAHAMAS           &  12    &   1.0  \\
                            &           &           & BAHAMAS-LowAGN    &  7.1    &   0.58  \\
                            &           &           & BAHAMAS-HighAGN   &  22     &   1.8  \\
                            &           &           & Horizon-AGN       &  2.8    &   0.23  \\
                            &           &           & IllustrisTNG100  &  1.9    &   0.15  \\
                            &           &           & IllustrisTNG300  &  2.0   &   0.16\\

            \end{tabular}
            \end{ruledtabular}
                        \caption{\label{tab:table2}Neutrino mass bias values. Models with stronger baryonic feedback have higher fractional $\sigma_{M_\nu}$ deviation from the fiducial value. See Fig.~\ref{fig:fig2}. }
        \end{table*}


\bibliography{references}

\end{document}